\documentclass{pasj00}

\begin{document}
\SetRunningHead{Hayakawa et al.}{Molecular and Atomic Gas toward HESS J1745-303}
\Received{2011/05/26}

\title{Molecular and Atomic Gas toward HESS J1745-303 in the Galactic Center: a Further Support for the Hadronic Scenario}

\author{Takahiro \textsc{Hayakawa}, %
  Kazufumi \textsc{Torii},
  Rei \textsc{Enokiya},
  Takanobu \textsc{Amano},
  and
  Yasuo \textsc{Fukui}
}
\affil{Department of Astrophysics, Nagoya University, Furo-cho, Chikusa-ku, Nagoya, 464-8602}
\email{t.hayakawa@a.phys.nagoya-u.ac.jp}


%

\KeyWords{ISM: clouds --- ISM: cosmic rays --- ISM: individual objects: HESS J1745-303} 

\maketitle

\begin{abstract}
We have compared the TeV $\gamma$-rays with the new $^{12}$CO $J$=2--1 data toward HESS J1745-303 in the Galactic center and confirmed that molecular cloud MG358.9-0.5 toward $(l, b)=(\timeform{358.9D}, \timeform{-0.5D})$ at $V_\mathrm{LSR}=-100$--$0$ km s$^{-1}$ shows a reasonable positional agreement with the primary peak (northern part) of the $\gamma$-ray source.
For the southern part of HESS J1745-303, we see no CO counterpart, whereas the H\emissiontype{I} gas in the SGPS H\emissiontype{I} dataset shows a possible counterpart to the $\gamma$-ray source.
This H\emissiontype{I} gas may be optically thick as supported by the H\emissiontype{I} line shape similar to the optically thick $^{12}$CO.
We estimate the total mass of interstellar protons including both the molecular and atomic gas to be $2\times 10^6 \MO$ and the cosmic-ray proton energy to be $6\times 10^{48}$ ergs in the hadronic scenario.
We discuss possible origins of the cosmic-ray protons including the nearby SNR G359.1-0.5.
The SNR may be able to explain the northern $\gamma$-ray source but the southern source seems to be too far to be energized by the SNR.
As an alternative, we argue that the second-order Fermi acceleration in the inter-clump space surrounded by randomly moving high-velocity clumps may offer a possible mechanism to accelerate protons.
The large turbulent motion with velocity dispersion of $\sim 15$ km s$^{-1}$ has energy density two orders of magnitude higher than in the solar vicinity and is viable as the energy source.
\end{abstract}

\section{Introduction}\label{sec:introduction}

Astronomical observations of $\gamma$-rays with space-borne and ground-based telescopes have become important tools to pursue physical processes in the Universe owing to the higher sensitivity and angular resolution than a decade ago.
In particular, TeV $\gamma$-ray observations with High-Energy Stereoscopic System (H.E.S.S.) have revealed more than 50 $\gamma$-ray sources in the Galactic plane, offering new tools to uncover the highest energy phenomena in the interstellar space (\cite{aharonian2006apj}; \yearcite{aharonian2006nature}).

Among the HESS sources, the extended source in the Galactic center (GC) is an outstanding object that shows a good spatial correlation with the dominant molecular feature, the central molecular zone (CMZ) \citep{morris1996}.
\citet{aharonian2006nature} noted the correlation between the molecular gas and TeV $\gamma$-rays by comparing CS \citep{tsuboi1999} and the TeV $\gamma$-ray image; in particular, \citet{aharonian2006nature} noted that three peaks of the extended $\gamma$-rays show a good spatial correlation with major molecular peaks in the CMZ, including the Sgr A, Sgr B2, and Sgr C molecular clouds (e.g., \cite{fukui1977}; \cite{scoville1975}; \cite{liszt1985}).
The good correlation suggests that the $\gamma$-rays are produced by the hadronic process by which TeV $\gamma$-rays are emitted via decay of neutral pions created in high energy reactions between the cosmic-ray protons and the interstellar protons.

HESS J1745-303 is another extended source on the negative longitude side of the GC at $(l, b)=(\timeform{358.9D}, \timeform{-0.5D})$.
This source seems to be separated from the CMZ by $1\degree$, while it is not clear if it is connected to the CMZ.
\citet{aharonian2008} studied this source and identified molecular gas observed at lower resolution of $9\arcmin$ as a candidate of the associated cloud but could not find any objects which can fully match the TeV $\gamma$-rays.
This molecular cloud (hereafter MG358.9-0.5) is located toward the Galactic-western edge of the shell-like distribution, which surrounds an SNR G359.1-0.5 \citep{uchida1992a}.
\citet{aharonian2008} searched for associated objects in the literature and listed up several candidates including pulsars and G359.1-0.5 while the origin of the cosmic-rays remained unclear.

We present here new $^{12}$CO $J$=2--1 observations with NANTEN2 and H\emissiontype{I} data obtained with the Parkes 64 m telescope toward HESS J1745-303.
We confirmed association of the suggested molecular cloud MG358.9-0.5 with the HESS source and identified other possible association between the HESS source and atomic gas.
Section \ref{sect:observations} gives observations and section \ref{sect:results} presents results.
Section \ref{sect:discussion} gives discussion and section \ref{sect:summary} summarizes the paper.

\section{Observations and Datasets}\label{sect:observations}
\subsection{$^\mathit{12}$CO $J$=2--1 Observations}

We made observations of the $^{12}$CO $J$=2--1 line (the rest frequency is 230.53800 GHz) using the NANTEN2 4m telescope at Atacama, Chile, in the period from July 2010 to January 2011.
The 4m-diameter provides a half power beam width (HPBW) of $90\arcsec$.
The receiver frontend was a 4 K cooled SIS receiver.
The spectrometer was a digital Fourier spectrometer (DFS) with a 1 GHz (corresponds to 1300 km s$^{-1}$ at 230 GHz) bandwidth and a 61 kHz (0.08 km s$^{-1}$) resolution.
The velocity $V_\mathrm{LSR}$ hereafter refers to the local standard of rest.
The pointing accuracy was estimated to be about $10\arcsec$ by observing Jupiter and Venus every two hours, and system stability was checked by observing M17SW ($\alpha=\timeform{10h20m24.4s}$, $\delta=\timeform{-16D13'18"}$ at J2000.0 coordinates) and Ori-KL ($\timeform{5h32m14.5s}$, $\timeform{-5D22'28"}$) every hour.
The standard chopper-wheel method was employed to calibrate intensity and to correct for atmospheric attenuation.
The adopted main beam efficiency $\eta_\mathrm{mb}$ was 0.5 as measured by observing Jupiter.

The observations were made as part of NANTEN2 Galactic-Center Survey which makes a mosaic of $15\arcmin \times 15\arcmin$ maps.
Each $15\arcmin$-square map was made by using on-the-fly (OTF) mapping technique.
The dump interval, the scan velocity, and the inter-scan distance were 1.0 sec, $30\arcsec$ per second, and $30\arcsec$ respectively, producing a Nyquist-sampled $30\arcsec$ grid map.
An off-source reference scan was obtained after every two lines of on-source scan.
The off-position of $(l, b)=(\timeform{1.133D}, \timeform{-1.467D}$) was checked to be free of emission with an r.m.s. noise level of 0.1 K for 61 kHz resolution.
In order to remove the scanning effect, the maps were obtained by scanning in both Galactic longitude and latitude directions, and combined using the algorithm described by \citet{emerson1998}.
The image is smoothed with a Gaussian kernel with FWHM of $60\arcsec$.

\subsection{$^\mathit{12}$CO $J$=1--0 Dataset}
We used the $^{12}$CO $J$=1--0 NANTEN Galactic Plane Survey (NGPS) dataset \citep{mizuno2004}.
The HPBW was $\timeform{2.6'}$ at 115 GHz.
The spectrometer was an acousto-optical spectrometer (AOS) with 2048 channels.
The frequency coverage and resolution were 250 MHz and 250 kHz, corresponding to velocity coverage of $\pm 300$ km s$^{-1}$ and a velocity resolution of 0.6 km s$^{-1}$, respectively.
The observations were carried out by position switching at a $4\arcmin$ grid spacing.
The on-source integration time per point was 4 s and the r.m.s. noise fluctuations of the spectral data were $\sim 0.3$ K at 0.6 km s$^{-1}$ velocity resolution.

\subsection{H\emissiontype{I} Dataset}

We also used the Southern Galactic Plane Survey (SGPS) dataset \citep{mccluregriffiths2005}.
The SGPS data were gridded with a cell size of $\timeform{4'}$ and a Gaussian smoothing kernel of $\timeform{16'}$.
The bandwidth and channel width of the spectrometer were 4.5 MHz and 3.9 kHz (950 and 0.82 km s$^{-1}$ at 1420 MHz), respectively.
For further details, see \citet{mccluregriffiths2005}.

\begin{figure}
  \begin{center}
    \FigureFile(80mm,80mm){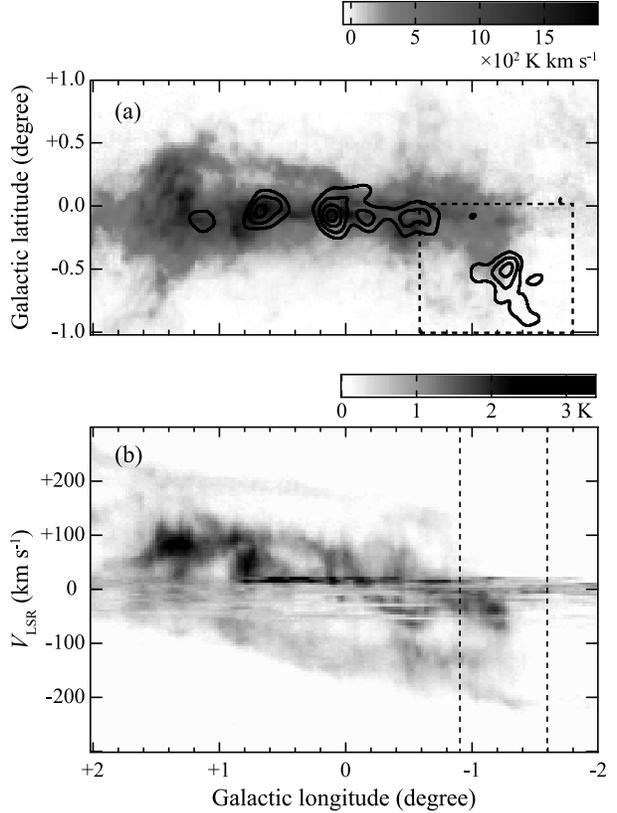}
  \end{center}
  \caption{
    (a) $^{12}$CO $J$=2--1 integrated-intensity map toward the Galactic center in a velocity range $20 \leq |V_\mathrm{LSR}| \leq 300$ km s$^{-1}$.
    The contours are TeV $\gamma$-ray count map (start at 300 counts and increment every 20 counts) after subtraction of the two dominant point sources Sgr A$^{\ast}$ and G0.9+0.1 \citep{aharonian2006nature}.
    The bounding box indicates an area of HESS J1745-303 shown in figure \ref{fig:intensitymap}.
    (b) A longitude-velocity diagram averaged over $-1\degree \leq b \leq 0\degree$.
    The vertical broken lines outline the position of HESS J1745-303 ($l=\timeform{358.4D}$ -- $\timeform{359.1D}$).
    }\label{fig:CMZ}
\end{figure}

\section{Results}\label{sect:results}
\subsection{CO Results}\label{sect:COresults}
\begin{figure}
  \begin{center}
    \FigureFile(80mm,160mm){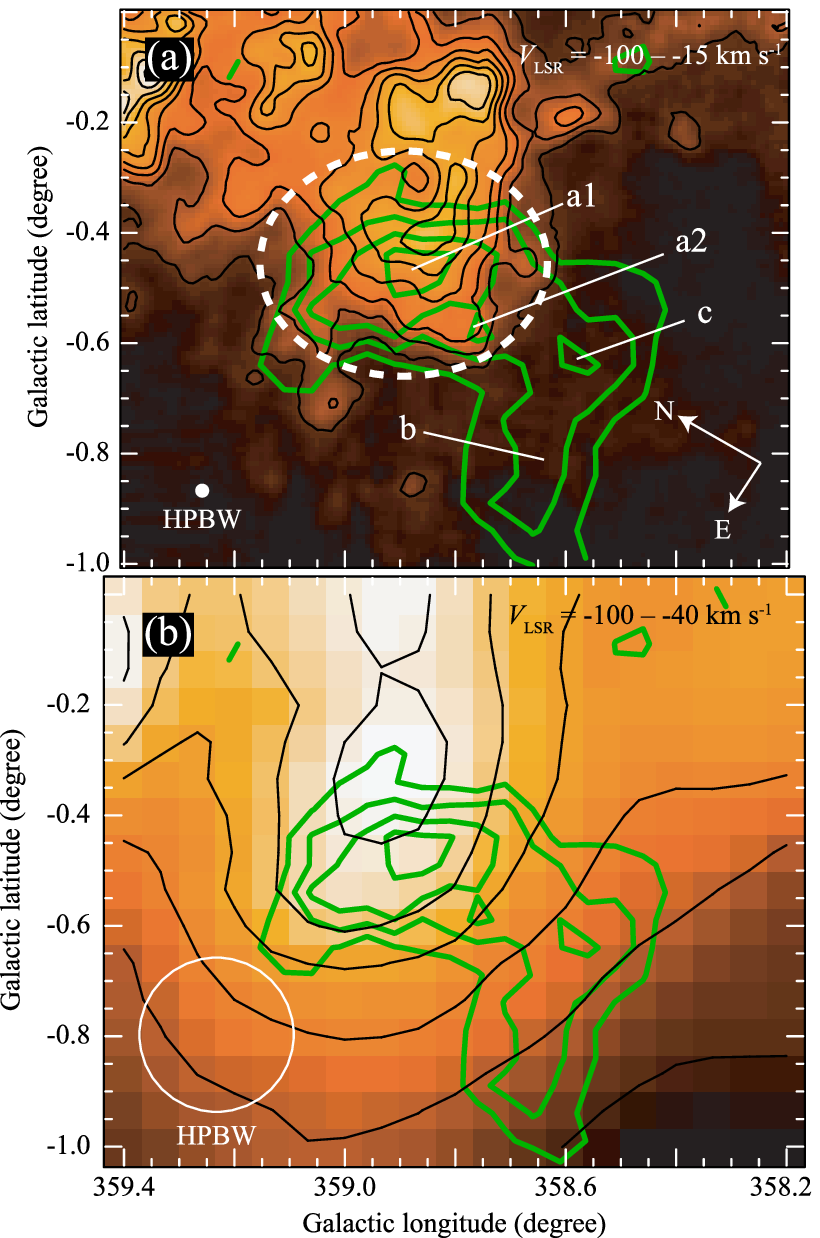}
  \end{center}
  \caption{
    (a) $^{12}$CO $J$=2--1 integrated-intensity map toward HESS J1745-303, overlaid with thin contours start at and increase in steps of 60 K km s$^{-1}$.
    The integration velocity range, from $-100$ to $-15$ km s$^{-1}$, is chosen to avoid contamination by local features in $V_\mathrm{LSR}\sim -15$ -- $+20$ km s$^{-1}$.
    The molecular cloud MG358.9-0.5 is outlined by the broken ellipse.
    Statistical significance map of HESS J1745-303 \citep{aharonian2008} is shown in thick contours (from $4\sigma$ to $7\sigma$).
    H\emissiontype{I} and $^{12}$CO profiles in figure \ref{fig:spectra} are toward four positions a1, a2, b, and c.
    The arrows indicate the directions of north and east.
    The HPBW indicated with a filled circle on the bottom-left corner is for $^{12}$CO data.
    (b) Integrated-intensity distribution of H\emissiontype{I} emission \citep{mccluregriffiths2005} overlaid with thin contours from 300 to 800 K km s$^{-1}$ with 100 K km s$^{-1}$ step.
    The integration velocity range is limited from $-100$ to $-40$ km s$^{-1}$ to exclude the local feature in $V_\mathrm{LSR}\gtrsim -40$ km s$^{-1}$, and subtracted the 3 kpc-arm component at $-60$ km s$^{-1}$.
    The HPBW indicated with an open circle is for H\emissiontype{I} data.
  }\label{fig:intensitymap}
\end{figure}

\begin{figure}
  \begin{center}
    \FigureFile(80mm,80mm){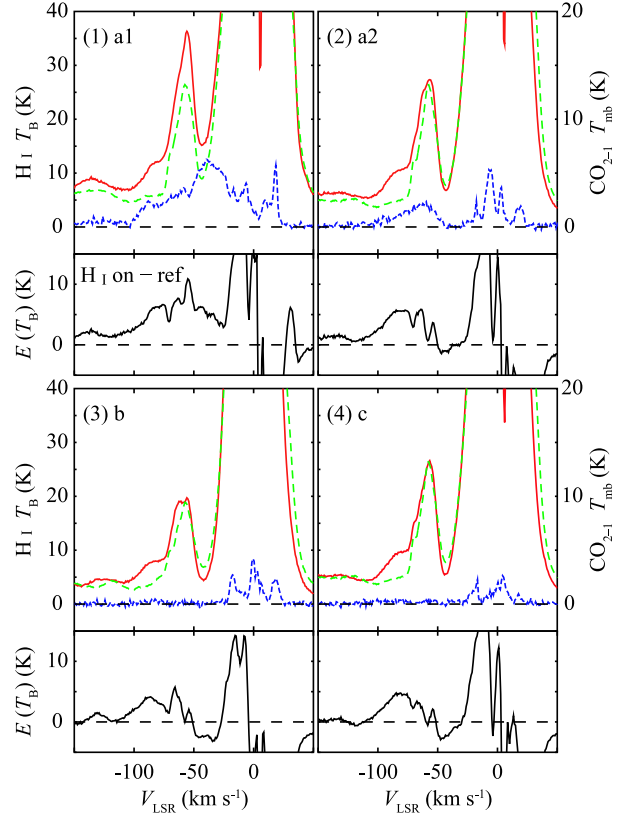}
  \end{center}
  \caption{
    Example of line profiles toward positions (1) a1, (2) a2, (3) b, and (4) c labeled in figure \ref{fig:intensitymap}.
    The solid and short-dashed lines on the upper side of each panel show H\emissiontype{I} and $^{12}$CO $J$=2--1 spectra, respectively.
    The long-dashed lines are reference H\emissiontype{I} spectra created by averaging two H\emissiontype{I} spectra outside HESS J1745-303 ($l=\timeform{358.2D}$ and $\timeform{359.4D}$) at the same $b$ with each on-HESS-source position.
    The subtracted H\emissiontype{I} spectrum is shown on the lower panel.
  }\label{fig:spectra}
\end{figure}

Figure \ref{fig:CMZ} shows the overall $^{12}$CO distribution in the ($l$, $b$) plane and velocity-$l$ plane toward the CMZ with the TeV $\gamma$-ray distribution overlaid.
We searched for possible counterparts in a velocity range from $-300$ to $+300$ km s$^{-1}$ in the $^{12}$CO $J$=2--1 dataset.
Figure \ref{fig:intensitymap}(a) shows an overlay of $^{12}$CO and TeV $\gamma$-rays in an area of HESS J1745-303.
The peak of MG358.9-0.5 is well coincident with the primary peak in the northern part of the TeV $\gamma$-rays, region A of \citet{aharonian2008}, as already noted by these authors.
We note that our new $^{12}$CO data have higher spatial resolution by a factor of 5 than the $^{12}$CO data used by \citet{aharonian2008}.
It seems that the cloud has velocities between $-100$ and around 0 km s$^{-1}$ toward the position a1 and $-100$ -- $-40$ km s$^{-1}$ toward the position a2 in region A (see table \ref{tab:1}), although the full velocity range of the molecular cloud is not well determined because of the contamination by the foreground emission at low velocities ($|V_\mathrm{LSR}|\lesssim 20$ km s$^{-1}$).
Sample spectra toward these two positions are shown in figure \ref{fig:spectra}.
The southern part of the $\gamma$-ray source (regions B and C of \cite{aharonian2008}), however, does not have any counterpart in $^{12}$CO.

\begin{table*}
  \caption{List of selected positions toward HESS J1745-303.}\label{tab:1}
  \begin{center}
    \begin{tabular}{lrrrrc@{}rrrrc}
      \hline 
      & \multicolumn{4}{c}{Position} & & \multicolumn{4}{c}{$^{12}$CO $J$=2--1\footnotemark[$*$]} & Region\footnotemark[$\dagger$]\\ \cline{2-5} \cline{7-10}
      & \multicolumn{1}{c}{$l$} & \multicolumn{1}{c}{$b$} & \multicolumn{1}{c}{$\alpha$ (J2000.0)} & \multicolumn{1}{c}{$\delta$ (J2000.0)} & & \multicolumn{1}{c}{$T_\mathrm{mb}$\footnotemark[$\ddagger$]} & \multicolumn{1}{c}{$V_\mathrm{c}$\footnotemark[$\S$]} & \multicolumn{1}{c}{$\Delta V$\footnotemark[$\|$]} & \multicolumn{1}{c}{$W$\footnotemark[$\#$]} & \\
      & \multicolumn{1}{c}{(degree)} & \multicolumn{1}{c}{(degree)} & & & & \multicolumn{1}{c}{(K)} & \multicolumn{1}{c}{(km s$^{-1}$)} & \multicolumn{1}{c}{(km s$^{-1}$)} & \multicolumn{1}{c}{(K km s$^{-1}$)} & \\
      \hline 
      a1 & 354.87 & $-0.47$ & $\timeform{17h34m43.9s}$ & $\timeform{-33D31'57"}$ & & 6.3 & $-57.2$ & 50.9 & 293.3 & A\\
      a2 & 354.73 & $-0.61$ & $\timeform{17h34m56.0s}$ & $\timeform{-33D43'34"}$ & & 2.4 & $-61.9$ & 33.2 & 82.4 & A\\
      b  & 354.67 & $-0.80$ & $\timeform{17h35m32.8s}$ & $\timeform{-33D52'45"}$ & & \multicolumn{1}{c}{---} & \multicolumn{1}{c}{---} & \multicolumn{1}{c}{---} & \multicolumn{1}{c}{---} & B\\
      c  & 354.60 & $-0.60$ & $\timeform{17h34m33.2s}$ & $\timeform{-33D49'48"}$ & & \multicolumn{1}{c}{---} & \multicolumn{1}{c}{---} & \multicolumn{1}{c}{---} & \multicolumn{1}{c}{---} & C\\
      \hline
      \multicolumn{11}{@{}l@{}}{\hbox to 0pt{\parbox{180mm}{\footnotesize
	    Notes.
	    \par\noindent
	    \footnotemark[$*$] Physical properties are analyzed in a velocity range between $-100$ and $-15$ km s$^{-1}$.
	    \par\noindent
	    \footnotemark[$\dagger$] Defined by \citet{aharonian2008} and shown in figure \ref{fig:gammaexcess}.

	    \footnotemark[$\ddagger$] Peak intensity.
	    \par\noindent
	    \footnotemark[$\S$] Centroid LSR velocity $V_\mathrm{c}=\int v T dv \big/ \int T dv$.
	    \par\noindent
	    \footnotemark[$\|$] Linewidth $\Delta V=2 \sqrt{2\ln2 \int (v-V_\mathrm{c})^2 T dv\big/\int T dv}$.
	    \par\noindent
	    \footnotemark[$\#$] Velocity integrated intensity $W=\int T dv$.
	    \par\noindent
	  }\hss}}
    \end{tabular}
  \end{center}
\end{table*}

MG358.9-0.5 shows $^{12}$CO line profiles peaked at high-velocity ($\sim -50$ km s$^{-1}$) with broad line-widths ($\sim 30$--$50$ km s$^{-1}$, FWHM) and absorption at $-60$ km s$^{-1}$ by the 3 kpc-arm component, as previously mentioned by \citet{uchida1992a}.
Thus, the cloud is not a local feature but in the GC perhaps near the CMZ.
Figure \ref{fig:CMZ} shows a strong hint that the cloud is physically linked to the CMZ from its continuity to the CMZ both in space and velocity, whereas such connection is not recognized in the CS image that does not cover $b$ below $\timeform{-0.25D}$.

The molecular cloud mass within a HESS $4\sigma$ significance contour is estimated to be $1.6\times 10^{6} \MO$ by adopting a distance of 8.5 kpc and a conversion factor between H$_2$ column density and $^{12}$CO integrated intensity in the GC, $X(^{12}\mathrm{CO}_{1-0})=N(\mathrm{H_2})/W(^{12}\mathrm{CO}_{1-0})=7.0\times 10^{19}$ cm$^{-2}$ (K km s$^{-1}$)$^{-1}$ \citep{torii2010}.
Here, we used $J$=1--0 integrated intensity instead of $J$=2--1, because the $N(\mathrm{H_2})/W(^{12}\mathrm{CO}_{2-1})$ conversion factor is not well established.
The $^{12}$CO $J$=1--0 integrated intensities of MG358.9-0.5 in a velocity range from $-100$ to $-15$ km s$^{-1}$ are $\sim 100$ -- 500 K km s$^{-1}$ and roughly proportional to those of $J$=2--1 with an intensity ratio of $W(^{12}\mathrm{CO}_{2-1})/W(^{12}\mathrm{CO}_{1-0})\sim 0.7$.
The average H$_2$ density is estimated to be $2\times 10^2$ cm$^{-3}$ as derived by dividing the mean H$_2$ column density $2\times 10^{22}$ cm$^{-2}$ by the projected size of the molecular cloud, 40 pc.

\subsection{H\emissiontype{I} Results}\label{sect:HIresults}

Figures \ref{fig:intensitymap}(b) shows H\emissiontype{I} distribution compiled from the H\emissiontype{I} dataset \citep{mccluregriffiths2005} in the same region.
Although the angular resolution $16\arcmin$ is not high enough, the H\emissiontype{I} distribution shows a possible counterpart to the TeV $\gamma$-rays.
Figure \ref{fig:spectra} shows H\emissiontype{I} and $^{12}$CO line profiles toward the four positions a1, a2, b and c (table \ref{tab:1}).

In order to test if the H\emissiontype{I} emission is similar to the $^{12}$CO emission, we made subtraction of the H\emissiontype{I} profiles toward the four positions and the reference H\emissiontype{I} profiles; the reference H\emissiontype{I} profiles for each were taken at the same $b$ separated by $\sim \timeform{0.5D}$ on the both sides ($l=\timeform{358.2D}$ and $\timeform{359.4D}$).
The results shown in figure \ref{fig:spectra} indicate that these subtracted H\emissiontype{I} profiles show fairly similar shapes and intensities to those of the $^{12}$CO profiles toward MG358.9-0.5 in their peak velocity and linewidth.
The similarity leads us to suggest that the H\emissiontype{I} emission is from H\emissiontype{I} gas physically connected to MG358.9-0.5 but without $^{12}$CO emission.
We note that the absorption of the radio continuum emission is not important in the direction (e.g., \cite{yusefzadeh2004}).

The H\emissiontype{I} emission in the negative velocity range consists of three features; the 3-kpc arm at $V_\mathrm{LSR}= -60$ km s$^{-1}$ \citep{vanwoerden1957}, the $V_\mathrm{LSR}=-100 \mbox{--} -75$ km s$^{-1}$ feature and the $V_\mathrm{LSR}\leq -100$ km s$^{-1}$ feature.
The 3-kpc arm is located in front of and outside the GC and is not seen in the $^{12}$CO emission in figure \ref{fig:spectra}.
The $V_\mathrm{LSR} \leq -100$ km s$^{-1}$ feature is likely distributed within 1 kpc of the GC  \citep{rougoor1960}.
The $-100 \mbox{--} -75$ km s$^{-1}$ feature is perhaps located within the inner few 100 pc, similar to the CMZ, and is physically related to the $^{12}$CO counterpart of HESS J1745-303, MG358.9-0.5.
The contamination by the 3-kpc arm masks the H\emissiontype{I} emission above $-75$ km s$^{-1}$, while the $^{12}$CO emission at the position a2 (figure \ref{fig:spectra}(2)) suggests the velocity of the H\emissiontype{I} counterpart may be extended to $-45$ km s$^{-1}$.

H\emissiontype{I} column density is calculated by the following relation if the line is optically thin,
\begin{equation}
N(\mathrm{H\emissiontype{I}})=1.823\times 10^{18} \int T_\mathrm{B}(v) dv\ \mbox{(cm$^{-2}$)}, 
\end{equation}
where $T_\mathrm{B}(v)$ is brightness temperature in unit of K.
The H\emissiontype{I} integrated intensities of 160--270 K km s$^{-1}$, which are given by assuming single-Gaussian profiles with FWHM line-widths of 30--50 km s$^{-1}$ and a peak intensity of 5 K, correspond to H\emissiontype{I} column density of $(3\mbox{--}5)\times 10^{20}$ cm$^{-2}$ when the line is optically thin as usually assumed.
The average H\emissiontype{I} density is then estimated to be 3--5 cm$^{-3}$ for a line-of-sight length of 30 pc.

An alternative is that the H\emissiontype{I} line may be optically thick, because the H\emissiontype{I} line profile seems fairly similar to that of the optically thick $^{12}$CO line (figure \ref{fig:spectra}).
If so, the H\emissiontype{I} emission is from dense atomic gas with lower temperature close to that of the $^{12}$CO cloud as suggested by the peak temperature of H\emissiontype{I} similar to $^{12}$CO; we are not able to estimate the H\emissiontype{I} density, but can constrain the average density to be $\lesssim 100$ cm$^{-3}$ by assuming that the H\emissiontype{I} is less dense to emit $^{12}$CO. 
Adopting spin temperature $T_\mathrm{s}\sim 30$--$60$ K which is comparable to the kinetic temperature of the molecular gas in the CMZ (e.g., \cite{gusten1981,morris1996,martin2004}), we estimate the H\emissiontype{I} column density to be $N(\mathrm{H\emissiontype{I}}) \gtsim (0.5\mbox{--}1)\times 10^{22}$ cm$^{-2}$ for the H\emissiontype{I} peak optical depth $\tau$ greater than $\sim 3$ and a linewidth of $\sim 30$ km s$^{-1}$ by using the following relation;

\begin{eqnarray}
N(\mbox{H\emissiontype{I}}) & = & \frac{32\pi \nu^2 k T_\mathrm{s}}{3 c^3 h A_\mathrm{ul}} \tau \Delta V \nonumber \\
& = & 1.823\times 10^{18} \tau \left(\frac{T_\mathrm{s}}{\mbox{K}}\right) \left(\frac{\Delta V}{\mbox{km s$^{-1}$}}\right)\ \mbox{(cm$^{-2}$)}
\end{eqnarray}
where $c$ is the light velocity, $h$ the Planck constant, $k$ the Boltzmann constant, $\nu=1420$ MHz the rest frequency and $A_\mathrm{ul}=2.85\times 10^{-15}$ s$^{-1}$ Einstein's $A$ coefficient.
We then obtain $n(\mathrm{H\emissiontype{I}}) \gtsim (0.5\mbox{--}1)\times 10^{2}$ cm$^{-3}$ by dividing the column density by the typical size of the $\gamma$-rays, $\sim 30$ pc.
Considering the upper limit for the average density, we infer that the density of the optically thick H\emissiontype{I} gas linked to MG358.9-0.5 is around 100 cm$^{-3}$.

\section{Discussion}\label{sect:discussion}

\begin{table*}
  \caption{Energy of cosmic-ray protons.}\label{tab:cosmicrayenergy}
  \begin{center}
    \begin{tabular}{lrrrr}
      \hline
      & \multicolumn{1}{c}{Full\footnotemark[$*$]} & \multicolumn{1}{c}{A} & \multicolumn{1}{c}{B} & \multicolumn{1}{c}{C} \\
      \hline
      $\Gamma$\footnotemark[$\dagger$] & 2.71 & 2.67 & 2.93 & 2.86 \\
      $w_\mathrm{\gamma}(\mbox{0.3--40 TeV})$\footnotemark[$\ddagger$] [$10^{-11}$ erg cm$^{-2}$ s$^{-1}$] & 1.46 & 0.22 & 0.11 & 0.14 \\
      $W_\mathrm{p}(\mbox{3--400 TeV})$\footnotemark[$\S$] [erg] \\
      ($\overline{n_\mathrm{p}}=10^2$ cm$^{-3}$) & $6\times 10^{48}$ & $9\times 10^{47}$ & $4\times 10^{47}$ & $5\times 10^{47}$ \\
      ($\overline{n_\mathrm{p}}=10$ cm$^{-3}$) & $6\times 10^{49}$ & $9\times 10^{48}$ & $4\times 10^{48}$ & $5\times 10^{48}$ \\
      \hline
      \multicolumn{4}{@{}l@{}}{\hbox to 0pt{\parbox{180mm}{\footnotesize
	    Notes.
	    \par\noindent
	    \footnotemark[$*$]{A region covering the entire HESS J1745-303.}
	    \par\noindent
	    \footnotemark[$\dagger$]{Photon index of TeV $\gamma$-rays \citep{aharonian2008}.}
	    \par\noindent
	    \footnotemark[$\ddagger$]{Energy flux in a energy range of 0.3--40 TeV, given by $\int E \phi(E) dE$, where $\phi(E)=dN/dE \propto (E/\mathrm{TeV})^{-\Gamma}$.}
	    \par\noindent
	    \footnotemark[$\S$]{Estimated total energy of cosmic-ray protons given by equation \ref{eqn:cosmicrayenergy}.}
	  }\hss}}
    \end{tabular}
  \end{center}
\end{table*}

The origin of the $\gamma$-rays in HESS J1745-303 has been discussed in the literature.
\citet{aharonian2008} reported that there is no significant spectral variability across HESS J1745-303 and suggested a single origin of the TeV $\gamma$-rays.
It is to be noted that the HESS detection is at $5\sigma$ level toward the southern part of HESS J1745-303 and the distribution of the spectral index is yet to be confirmed by higher sensitivity.
The authors also showed that the $\gamma$-ray source has no extended X-ray source in the {\it XMM-Newton} data.
\citet{bamba2009} confirmed the lack of non-thermal X-ray counterpart and argued that the leptonic model requires weak magnetic field strength, $B\lesssim 6$ $\mu$G in a $n=0.1$ cm$^{-3}$ model and $B \lesssim 10^2$ $\mu$G in a $n=5\times 10^3$ cm$^{-3}$ model, to explain an observed upper limit of X-ray flux in the $2$--$10$ keV band, $2.1\times 10^{-13}$ erg s$^{-1}$ cm$^{-2}$.
It is, on the other hand, shown that the field strength is at least 50 $\mu$G over 400 pc in the CMZ where the average gas density is $10^2$--$10^3$ cm$^{-3}$ \citep{crocker2010} and such strong magnetic field seems to be inconsistent with the absence of the X-rays.
This suggests that the hadronic scenario is favorable in HESS J1745-303.
The positional coincidence with the unidentified EGRET source 3EG J1744-3011 \citep{hartman1999}, also known as the AGILE first-catalog source 1AGL J1746-3017 \citep{pittori2009}, again supports the hadronic scenario which usually predict bright GeV $\gamma$-rays, though positional accuracy of the EGRET/AGILE source is not high enough.

\citet{aharonian2008} argued for the hadronic process that the $^{12}$CO cloud MG358.9-0.5 is responsible for the northern part of HESS J1745-303, whereas the ISM corresponding to the southern part of HESS J1745-303 remained unidentified.
We argue that the $^{12}$CO cloud has a dense H\emissiontype{I} envelope with no $^{12}$CO emission and that the H\emissiontype{I} envelope is responsible for the $\gamma$-ray production by cosmic-ray protons; it is likely that the H\emissiontype{I} envelope has moderately high density like $\sim 100$ cm$^{-3}$ and $T_\mathrm{s}$ of 30--60 K, significantly less than $\sim 100$ K (see section \ref{sect:HIresults}).
We recall that the Galactic $\gamma$-ray emission requires the component unseen either in $^{12}$CO or warm H\emissiontype{I} as indicated by an analysis of EGRET data \citep{grenier2005}.
The ``dark gas'' component in $^{12}$CO or H\emissiontype{I} in a Galactic scale is also recognized in the visual/infrared extinction (e.g., \cite{dobashi2011}) and in the dust emission obtained by the Planck satellite \citep{planck2011}.
In the TeV-$\gamma$-ray SNR RXJ1713.7-3946, it has been shown that cold H\emissiontype{I} gas observed as self-absorption dips matches well with the TeV-$\gamma$-ray shell, suggesting that such H\emissiontype{I} having density $\sim 100$ cm$^{-3}$ with no $^{12}$CO emission works as target protons in the hadronic interaction \citep{fukui2011}.
Considering these, we suggest that the H\emissiontype{I} cloud with no $^{12}$CO emission is a good candidate for the TeV $\gamma$-rays in the southern part and the TeV $\gamma$-ray emission of HESS J1745-303 is likely due to accelerated protons interacting with the molecular/atomic clouds in the GC region.

\begin{figure}
  \begin{center}
    \FigureFile(80mm,80mm){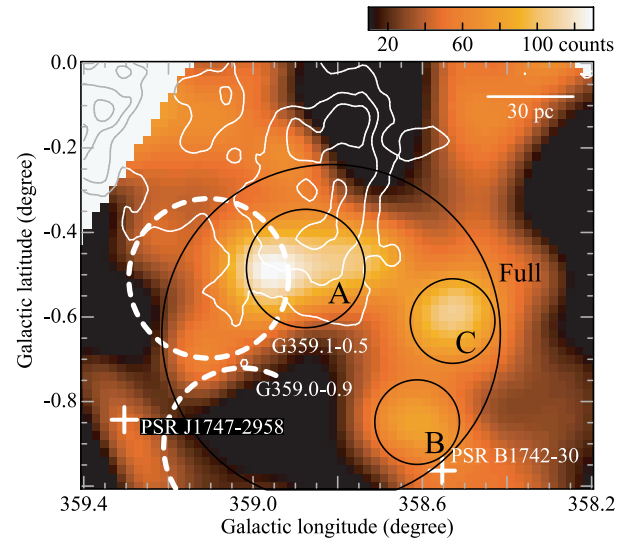}
  \end{center}
  \caption{
    TeV $\gamma$-ray excess count image overlaid with the $^{12}$CO $J$=2--1 integrated-intensity distribution (thin contours start at and increase in steps of 120 K km s$^{-1}$).
    Thin circles represent regions A, B, C and Full (a region covering the entire HESS J1745-303) of \citet{aharonian2008}, whose radii are $\timeform{0.14D}$, $\timeform{0.1D}$, $\timeform{0.1D}$ and $\timeform{0.4D}$, respectively.
    The crosses show the positions of pulsars PSR B1742-30 and J1747-2958.
    The broken thick circle and arc outline radio shells G359.1-0.5 and G359.0-0.9, respectively.
  }\label{fig:gammaexcess}
\end{figure}

Figure \ref{fig:gammaexcess} indicates the positions of nearby candidates for the high-energy cosmic-ray proton injector.
Among them, pulsars PSR B1742-30 and J1747-2958 and a SNR G359.0-0.9 are at heliocentric-distances of 2.08, 2.49, and 6 kpc respectively \citep{manchester2005, bamba2000} and, thus, unrelated to HESS J1745-303, whereas G359.1-0.5 is likely to be at a distance of the GC (e.g., \cite{uchida1992b}).
We discuss whether or not G359.1-0.5 is responsible for the cosmic-ray protons that produce the $\gamma$-rays in HESS J1745-303.
Adopting a distance of 8.5 kpc, the total energy of cosmic-ray protons required to generate the observed TeV $\gamma$-ray flux can be estimated by the following relation (e.g., \cite{aharonian2006aap}, \yearcite{aharonian2007});
\begin{eqnarray}\label{eqn:cosmicrayenergy}
W_\mathrm{p} & \simeq & t_\mathrm{pp\rightarrow \pi^0}\cdot L_\mathrm{\gamma} \nonumber \\
& \simeq & 3.9\times 10^{50} \left(\frac{\overline{n_\mathrm{p}}}{1\,\mathrm{cm}^{-3}}\right)^{-1} \left(\frac{w_\mathrm{\gamma}}{10^{-11}\,\mbox{erg cm$^{-2}$ s$^{-1}$}}\right)\,\mbox{erg},
\end{eqnarray}
where $t_\mathrm{pp\rightarrow \pi^0}\simeq 4.5\times 10^{15} (\overline{n_\mathrm{p}}/1\,\mathrm{cm}^{-3})^{-1}$ s is the characteristic cooling time of protons through the $\pi^0$ production channel, $L_\mathrm{\gamma}=4\pi (\mbox{8.5 kpc})^2 w_\mathrm{\gamma}$ the $\gamma$-ray luminosity, $\overline{n_\mathrm{p}}$ the mean proton density of the target ISM, and $w_\mathrm{\gamma}$ the $\gamma$-ray energy flux.
Table \ref{tab:cosmicrayenergy} summarizes the $w_\mathrm{\gamma}$ in an energy range of 0.3--40 TeV and estimated $W_\mathrm{p}$ in the corresponding energy range approximately 3--400 TeV for regions A, B, C and a region covering the entire HESS J1745-303 (named ``Full'' in \cite{aharonian2008}).

Region A is close to G359.1-0.5, while regions B and C are $\timeform{0.5D}$ far away from it.
The solid angle of region A as viewed from the center of G359.1-0.5 is about an order of magnitude larger than those of regions B and C.
The number flux of the cosmic-ray protons from the SNR should depend on distance to the target clouds, whereas the difference among the observed TeV $\gamma$-ray flux is not so significant.
Therefore, observed $\gamma$-ray flux from HESS J1745-303 does not support that the cosmic-ray protons originate from the SNR shell.

Alternatively, we suggest that a more spatially distributed origin of cosmic ray protons over the $\gamma$-ray source could be working in explaining the origin of HESS J1745-303.
It is suggested that particle acceleration throughout the inter-cloud medium is responsible for the cosmic rays in the CMZ (e.g., \cite{melia2011}).
The dense gas clumps are moving at a high velocity dispersion of 15 km s$^{-1}$ typical to the CMZ and in the other GC molecular clouds (e.g., \cite{morris1996}, \cite{fukui2006}).
The present work showed that the beam-filling factor of the $^{12}$CO and dense H\emissiontype{I} emitting clumps is 0.1--0.3, indicating that there exits low-density inter-clump space.
The average density of the $^{12}$CO and H\emissiontype{I} clumps is in the order of 100 cm$^{-3}$, not much different from each other within a factor of a few (see section \ref{sect:results}), and that of the inter-clump medium is likely less than 10 cm$^{-3}$.
We argue that the second-order Fermi acceleration is taking place in the low-density inter-clump space, which is common between the $^{12}$CO and H\emissiontype{I} clumps, and the accelerated cosmic-ray protons interact with the clumps to produce hadronic $\gamma$-rays via neutral pion decay.
The $\gamma$-rays may therefore have a similar spectral index over the entire sources as suggested \citet{aharonian2006apj} since both the $^{12}$CO and dense H\emissiontype{I} gas share similar inter-clump properties.
We note the energy density of the large turbulent motion in the GC is about 100 times higher than in the solar vicinity and is a reasonable source for the particle acceleration.
We note that the three major peaks of TeV $\gamma$-rays in the CMZ coincide with the molecular peaks with enhanced velocity dispersion (see figure \ref{fig:CMZ}), as is consistent with the second-order Fermi acceleration.
Amano \etal\ (2011, in preparation) have studied the second-order Fermi acceleration in the CMZ based on acceleration/energy-loss time scales and showed that the second-order Fermi acceleration is able to accelerate protons up to $\sim 100$ TeV range at density less than 10 cm$^{-3}$.
It is important to pursue this possibility further in order to better understand the $\gamma$-rays in the GC by exploring more the role of the second-order Fermi acceleration in the central few 100 pc.

\section{Summary}\label{sect:summary}

We have used the new $^{12}$CO $J$=2--1 data and the SGPS H\emissiontype{I} data to search for counterparts of the extended TeV $\gamma$-ray source HESS J1745-303 in the GC.
We suggest that the hadronic scenario is more favorable than the leptonic scenario because the fairly large magnetic field of more than 50 $\mu$G is unfavorable to cosmic ray electrons to produce $\gamma$-rays.
The molecular cloud MG358.9-0.5 shows a reasonable positional agreement with the northern part of HESS J1745-303 as already noted by \citet{aharonian2008}.
For the southern part of TeV $\gamma$-rays, we see no $^{12}$CO counterpart, whereas the H\emissiontype{I} gas with kinematics similar to the molecular cloud is a possible counterpart.
We confirmed that the H\emissiontype{I} gas is not local features but in the GC perhaps near the CMZ as supported by their continuity to the CMZ both in space and velocity.
We discuss possible origins of the cosmic-ray protons and found that the SNR G359.1-0.5 is not likely the source of cosmic-ray protons in the hadronic scenario because the estimated total energy of cosmic-ray protons required to produce the observed TeV $\gamma$-ray flux is more than what can be supplied by the SNR.
We argue that spatially extended injection of cosmic-ray protons is more plausible and suggest that the second-order Fermi acceleration offers another possible mechanism to accelerate cosmic-ray protons.
The turbulent motion with a velocity span of 30 -- 100 km s$^{-1}$ commonly seen in the CMZ including MG358.9-0.5 can offer a sufficiently large energy supply in this picture.

\bigskip

We thank the all members of the NANTEN2 consortium for the operation and persistent efforts to improve the telescopes.
The NANTEN project is based on a mutual agreement between Nagoya University and the Carnegie Institution of Washington (CIW).
We greatly appreciate the hospitality of all the staff members of the Las Campanas Observatory of CIW.
We are thankful to many Japanese public donors and companies who contributed to the realization of the project.
This work is financially supported in part by Grant-in-Aid for Scientific Research from JSPS (Core-to-core program 17004) and the Global COE Program of Nagoya University ``Quest for Fundamentals Priniciples in the Universe (QFPU)'' from JSPS and MEXT of Japan.




\begin{thebibliography}{}
\bibitem[Ade et al.(2011)]{planck2011}
  Ade,~P.~A.~R., \etal\ 2011, arXiv:1101.2029
\bibitem[Aharonian et al.(2006a)]{aharonian2006apj}
  Aharonian,~F., \etal\ 2006a, \apj, 636, 777
\bibitem[Aharonian et al.(2006b)]{aharonian2006nature}
  Aharonian,~F., \etal\ 2006b, \nat, 439, 695
\bibitem[Aharonian et al.(2006c)]{aharonian2006aap}
  Aharonian,~F., \etal\ 2006c, \aap, 449, 223
\bibitem[Aharonian et al.(2007)]{aharonian2007}
  Aharonian,~F., \etal\ 2007, \apj, 661, 236
\bibitem[Aharonian et al.(2008)]{aharonian2008}
  Aharonian,~F., \etal\ 2008, \aap, 483, 509
\bibitem[Bamba et al.(2000)]{bamba2000}
  Bamba,~A., Yokogawa,~J., Sakano,~M., \& Koyama,~K. 2000, \pasj, 52, 259
\bibitem[Bamba et al.(2009)]{bamba2009}
  Bamba,~A., Yamazaki,~R., Kohri,~K., Matsumoto,~H., Wagner,~S., P\"uhlhofer,~G., \& Kosack,~K. 2009, \apj, 691, 1854
\bibitem[Crocker et al.(2010)]{crocker2010}
  Crocker,~R.~M., Jones,~D.~I, Melia,~F., Ott,~J., \& Protheroe,~R.~J. 2010, \nat, 463, 65 
\bibitem[Dobashi(2011)]{dobashi2011}
  Dobashi,~K. 2011, \pasj, 63, 1
\bibitem[Emerson \& Gr\"aver(1998)]{emerson1998}
  Emerson,~D.~T., \& Gr\"ave,~R. 1998, \aap, 190, 353
\bibitem[Fukui et al.(1977)]{fukui1977}
  Fukui,~Y., Iguchi,~T., Kaifu,~N., Chikada,~Y., Morimoto,~M., Nagane,~K., Miyazawa,~K., \& Miyaji,~T. 1977, \pasj, 29, 643
\bibitem[Fukui et al.(2006)]{fukui2006}
  Fukui,~Y., \etal\ 2006, Science, 314, 106
\bibitem[Fukui et al.(2011)]{fukui2011}
  Fukui,~Y., \etal\ 2011, arXiv: 1107.0508
\bibitem[Grenier et al.(2005)]{grenier2005}
  Grenier,~I.~A., Casandjian,~J., \& Terrier,~R. 2005, Science, 307, 1292 
\bibitem[G\"usten et al.(1981)]{gusten1981}
  G\"usten,~R., Walmsley,~C.~M., \& Pauls,~T., 1981, 103, 197
\bibitem[Hartman et al.(1999)]{hartman1999}
  Hartman,~R.~C., \etal\ 1999, \apjs, 123, 79
\bibitem[Liszt(1985)]{liszt1985}
  Liszt,~H.~S. 1985, \apj, 293, L65
\bibitem[Manchester et al.(2005)]{manchester2005}
  Manchester,~R.~N., Hobbs,~G.~B., Teoh,~A., \& Hobbs,~M. 2005, \apj, 129, 1993
\bibitem[Martin et al.(2004)]{martin2004}
  Martin,~C.~L., Walsh,~W.~M., Xiao,~K., Lane,~A.~P., Wlaker,~C.~K., \& Stark,~A.~A., 2004, \apjs, 150, 239
\bibitem[McClure-Griffiths et al.(2005)]{mccluregriffiths2005}
  McClure-Griffiths,~N.~M., Dickey,~J.~M., Gaensler,~B.~M., Green,~A.~J., Haverkorn,~M., \& Strasser,~S. 2005, \apjs, 158, 178
\bibitem[Melia and Fatuzzo(2011)]{melia2011}
  Melia,~F., \& Fatuzzo,~M. 2011, \mnras, 410, L23
\bibitem[Mizuno \& Fukui(2004)]{mizuno2004}
  Mizuno,~A., \& Fukui,~Y. 2004, ASP Conf. Ser., 317, 59
\bibitem[Morris \& Serabyn(1996)]{morris1996}
  Morris,~M., \& Serabyn,~E. 1996, \araa, 34, 645
\bibitem[Pittori et al.(2009)]{pittori2009}
  Pittori,~C., \etal\ 2009, \aap, 506, 1563
\bibitem[Rougoor \& Oort(1960)]{rougoor1960}
  Rougoor,~G.,~W., Oort,~J.,~H., 1960, PNAS, 46, 1
\bibitem[Scoville et al.(1975)]{scoville1975}
  Scoville,~N.~Z., Solomon,~P.~M., \& Penzias,~A.~A. 1975, \apj, 201, 352
\bibitem[Torii et al.(2010)]{torii2010}
  Torii, K., et al. 2010, \pasj, 62, 1307 
\bibitem[Tsuboi et al.(1999)]{tsuboi1999}
  Tsuboi,~M., Handa,~T., \& Ukita,~N. 1999, \apj, 120, 1
\bibitem[Uchida et al.(1992a)]{uchida1992a}
  Uchida,~K., Morris,~M., Bally,~J., Pound,~M., \& Yusef-Zadeh,~F. 1992a, \apj, 398, 128
\bibitem[Uchida et al.(1992b)]{uchida1992b}
  Uchida,~K., Morris,~M., \& Yusef-Zadeh,~F. 1992b, \aj, 104, 533
\bibitem[van Woerden et al.(1957)]{vanwoerden1957}
  van Woerden,~H., Rougoor,~G.~W., \& Oort,~J.~H., 1957, CRAS, 244, 1691
\bibitem[Yusef-Zadeh et al.(2004)]{yusefzadeh2004}
  Yusef-Zadeh,~F., Hewitt,~J.~W. \& Cotton,~W., 2004, \apjs, 155, 421
\end{thebibliography}
\end{document}